\documentstyle[prl,aps,epsf,multicol]{revtex}

\newcommand{\be}{\begin{equation}}
\newcommand{\ee}{\end{equation}}
\newcommand{\bea}{\begin{eqnarray}}
\newcommand{\eea}{\end{eqnarray}}

\def\figone#1#2#3{\begin{figure}[!ht]
\centering \leavevmode
\epsfxsize=0.8\columnwidth \epsfbox{#1}
\caption{#2 \label{#3}}
\end{figure} }

\begin{document}

\title{Trap dominated dynamics of classical dimer models} 
\author{Dibyendu Das, Jan\'e Kondev and Bulbul Chakraborty}
\address{Martin Fisher School of Physics, Brandeis University, 
Mailstop 057, Waltham, Massachusetts 02454-9110, USA}
\maketitle

\begin{abstract}

We consider dynamics of  classical dimer models undergoing a phase 
transition to an ordered, frozen state. Relaxation processes  are dominated by traps 
which are entropic in origin and can be traced to the locally jammed nature 
of the dimer states. Depending on the nature of the phase transition,  critical dynamics 
are characterized either by an exponential, or sub-exponential in time  relaxation of the order parameter.
In the latter case relaxation time scales diverge {\em exponentially} as the critical point is approached,  
following the Vogel-Fulcher law. 

\end{abstract}


\begin{multicols}{2}

A variety of systems such as supercooled liquids, colloids, granular
matter and foams,  exhibit a transition from a flowing fluid phase
to a frozen solid phase. Jamming due to spatial constraints imposed 
on the elementary constituents of these materials 
has been proposed as a possible common cause of this dynamical
arrest \cite{edwards,nagel_jamming,weitz}. 
Model systems, such as hard spheres, have an important role to play in
the investigation of such a scenario since they allow for a precise
definition of jamming \cite{torquato}. They are also useful in
elucidating the precise relationship between thermodynamics and
dynamics in materials exhibiting a jammed phase \cite{krauth}. 
%Renewed
%interest in these long standing questions has been sparked by
%observations of spatially heterogeneous dynamics in polymer films
%\cite{israeloff} and colloids \cite{weitz} near the glass transition.

Dimer models are examples of jammed systems which have the 
added advantage of being exactly solvable \cite{nagle_review}. 
States of the dimer model are  specified by placing 
dimers on the bonds of the lattice so that every site is covered by 
exactly one dimer; see Fig~\ref{dimer_fig}. These dimer coverings are 
``locally jammed'' \cite{torquato} as each dimer cannot move to an empty neighboring bond 
without violating the packing constraint. Moves that involve {\em loops} of dimers 
and adjacent empty bonds, on the other hand, are allowed.  
An example of such a move for the hexagonal lattice involving an elementary plaquette is  
shown in Fig.~\ref{dimer_fig}a.  (Stochastic dynamics of the dimer model on the square lattice
based on these elementary moves were considered by Henley \cite{clh}.)
Most states of the dimer model allow for elementary moves; an example  of one 
which does not is shown in  Fig.~\ref{dimer_fig}b. The smallest change in this case involves a
system spanning loop, and we call this state ``maximally  jammed''. 
If we define an energy functional on the space of dimer coverings which favors the maximally jammed state, 
a transition into this state will occur as we lower the temperature.  
The central question we address in this letter is: {\em What happens to relaxation time scales of the dimer model 
as the transition to the maximally jammed state is approached?} 
We will show that the relaxation is dominated by entropy barriers and is sensitive to the  
thermodynamics of the model near the phase transition point. 

\figone{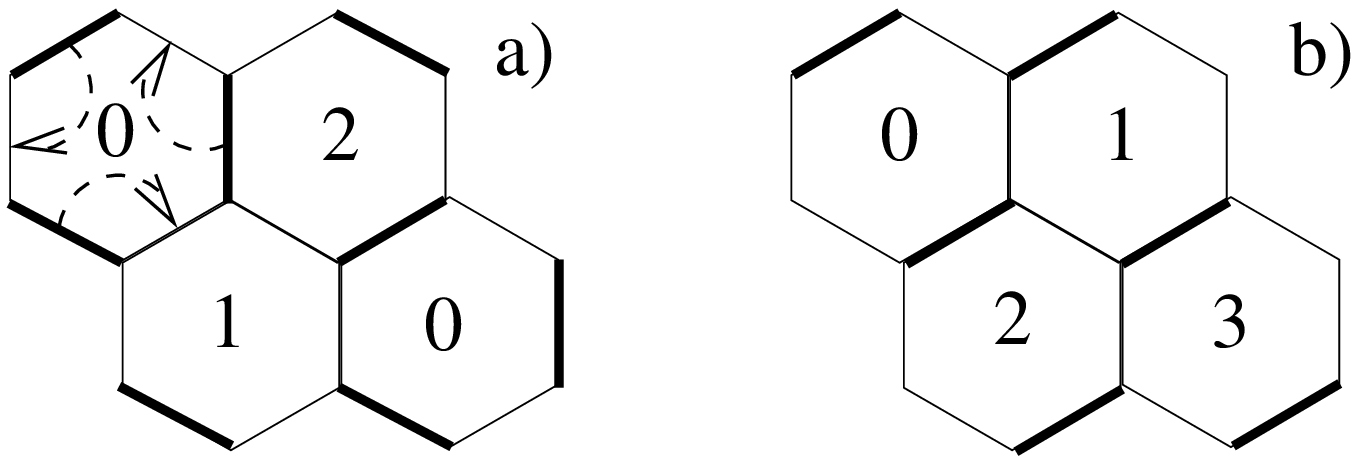}{(a) Dimer covering of the honeycomb lattice with an elementary loop update  
indicated by the arrows. The numbers are the heights of the equivalent interface.  (b) An ordered, maximally 
jammed dimer covering; the equivalent interface is tilted with maximum slope.}{dimer_fig}

We consider two energy functionals. One leads to a Pokrovsky-Talapov type transition 
to the maximally jammed state \cite{nagle_review}. The other exhibits a continuous transition to the same state 
along a metastable line, and  belongs to  a different universality class. 
For both transitions we find strong departures from 
the canonical critical-slowing-down scenarios \cite{halperin} 
due to the presence of entropy barriers. Barriers can be traced directly 
to the non-local nature of the dynamical moves allowed by the jammed states. 
For the transition that occurs along a metastable line, time scales associated with 
critical fluctuations diverge exponentially following a Vogel-Fulcher-like form. This 
is reminiscent of what is observed in fragile glass formers \cite{angel_review}.   
To our knowledge this is the first example of a system with no quenched in disorder 
which exhibits this type of critical dynamics.

\paragraph{Dimer models} 
We consider the  dimer model on the  2-d hexagonal
lattice of linear size $L$, having $2 L^2$ sites and $3 L^2$ bonds, 
with periodic boundary conditions \cite{kasteleyn}. 
A useful representation of the dimer model is given by the height map which 
associates a discrete interface $h(x,y)$ with every dimer covering \cite{blote}. 
The heights of the interface are defined on the vertices of the dual 
triangular lattice. The height difference $\Delta$ between two nearest neighboring 
sites is -2 or +1 depending on whether the bond of the honeycomb lattice 
that separates them is occupied by a dimer or not; see Fig.~\ref{dimer_fig}a. 
Directions in which the height change is $+\Delta$ are specified by orienting 
all the up pointing triangles of the dual lattice clockwise. 

The dimer model has an extensive entropy.  The ensemble of equal
weighted dimer coverings maps to a rough surface with a
gradient-square free energy \cite{blote}. Fluctuations of the surface
are entropic in origin. A phase transition can be induced in the dimer
model by including an energy functional which is minimized by a dimer
covering corresponding to a smooth, {\em maximally tilted} surface
which corresponds to the maximally jammed state shown in
Fig.~\ref{dimer_fig}b.  Here we discuss two such energy functionals.

For periodic boundary conditions the tilt vector, $({\Delta}_x h ,
{\Delta}_y h)$, where ${\Delta}_{x,y} h$ is the average height difference
in the $x$ or $y$ direction, has only one independent component
$\rho$ \cite{foot2}.  In terms of $\rho$, the energy functionals can be
written as:
\be E({\rho}) = \left \{ \begin{array}{ll}
-({{\mu L^2} / 3})~~({1}+{8}{{\rho}}^2) & {(I)}  
\\ -({{\mu L^2} / 3})~~(1+2{{\rho}}) & {(K)}  
\end{array} \right. ; 
\label{energyF}
\ee
$\mu$ is a dimensionless coupling that drives the transition. 

The entropy of the dimer model as a function of $\rho$ 
was calculated exactly \cite{kasteleyn,adhar}: 
\be
S({\rho}) = L^2 \left \{ {{2 \ln2} \over 3} (1-{\rho}) +  
{2 \over \pi}\int_{0}^{{\pi \over 3}(1-{\rho})}dx \ln[\cos x] \right \}
\label{entropyD1}
\ee 
This function has a maximum at $\rho=0$ which is the equilibrium value at
$\mu=0$. For finite $\mu$ the $I$ model was previously considered in
Ref.~\cite{hui}, while the $K$ model was solved by Kasteleyn
\cite{kasteleyn}.

In the $I$ model, with the free energy  $F=E-S$, and  the energy and
entropy given by Eqs.~\ref{energyF} and~\ref{entropyD1}, there is an
interesting phase transition along the metastable line,
when the order parameter is confined to the free energy well around
the zero-tilt state.  Namely, at $\mu_{*} = \pi/(8\sqrt{3})$ (the
end-point of the metastable line) the order parameter ${\rho}$ has a
discontinuous jump from $0$ to $1$, characteristic of a first-order
transition.  At the same time, as $\mu_{*}$ is approached from below,
fluctuations of ${\rho}$ around $0$ diverge, as would be expected at a
critical point.  This transition was discussed in detail in Ref.~\cite{hui}.

By contrast, in the $K$ model,  there is a continuous variation of
the equilibrium value of the order parameter, ${{\rho}}_{min}$, 
with $\mu$. On varying $\mu$ from $0$ to the transition
point $\mu_{*} = {\rm ln}2$, ${{\rho}}_{min}$ varies from $0$ to
$1$. Apart from the shift of the free energy minimum, the fluctuations
also grow with increasing $\mu$, and at the transition, the curvature
of the free energy at its  minimum goes to zero.

\paragraph{Coarse-grained dynamics} 
As mentioned in the introduction, the hard constraint of
non-overlapping of dimers, gives rise to nonlocal dynamics. We consider  
Monte-Carlo dynamics based on loop updates with 
loops of arbitrary size; a concrete implementation 
is given in Ref.~\cite{krauth1}. Since  we
take periodic boundary conditions, loops with different winding
numbers can be formed. We restrict loop updates to loops with winding
numbers $(0,0), (1,0)$ and $(0,1)$, only.  The microscopic transition rates for
loop updates are given by Metropolis rules that follow from
Eq.~\ref{energyF}.

Given the microscopic loop dynamics, which satisfy conditions of ergodicity and 
detailed balance, we ask what are the coarse-grained dynamics of the order 
parameter, $\rho$.  Since the energy functions in Eq.~\ref{energyF} depend on the global tilt
$\rho$ only, it follows that all updates of topologically trivial
loops (i.e.~those with $(0,0)$ winding number) have $\Delta E =
0$. Only when system spanning loops with nonzero winding numbers are
updated does the energy of the state change \cite{foot3}. 
This feature naturally leads to fast and slow processes in the
Monte-Carlo dynamics. On a faster time scale, non-winding loops are
updated with no effect on the overall tilt of the surface, while on a
much slower time scale, winding loops are updated causing a change
in the tilt of the surface.

The {\it coarse-grained} dynamics of global tilt changes are described 
by a  master equation, which features an  unusual form for the transition 
rates between different tilt states. Namely, the rates of transitions from 
higher into lower tilt states are determined by the energy barrier alone:
\be 
W_{{\rho-1/L}, \rho} = \Gamma_o e^{-(E(\rho-1/L)-E(\rho))} \ ; 
\label{barr1}
\ee
here $\Gamma_o$ is a constant. 
This follows from the observation that in order to lower the tilt and increase the 
energy an existing system spanning loop needs to be updated. 

{}From detailed balance, 
$W_{{\rho-1/L}, \rho}/ W_{{\rho},{\rho-1/L}} = \exp[-(F(\rho-1/L)-F(\rho))]$,  
we conclude that the rates of transitions to higher tilt states must be determined by the entropy: 
\be
W_{{\rho},{\rho-1/L}} = \Gamma_o  e^{-(S(\rho-1/L)-S(\rho))} \ .
\label{barr2}
\ee
Eq.~\ref{barr2} can also be argued from the following observation:  
to increase the tilt and lower the energy, a new system-spanning loop
has to be accommodated by (typically) many rearrangements of the
topologically trivial loops. Once a new system spanning loop is
introduced, and a higher tilt state is reached, the entropy decreases.
The form of the transition rates that we are arguing for here, was
directly observed in numerical simulations of the three coloring model
\cite{loops_epl}, which is a close relative of the dimer model. The
two are equivalent if, in the dimer models, a weight of 2 is attached
to each loop formed by bonds that are not covered by dimers.

\figone{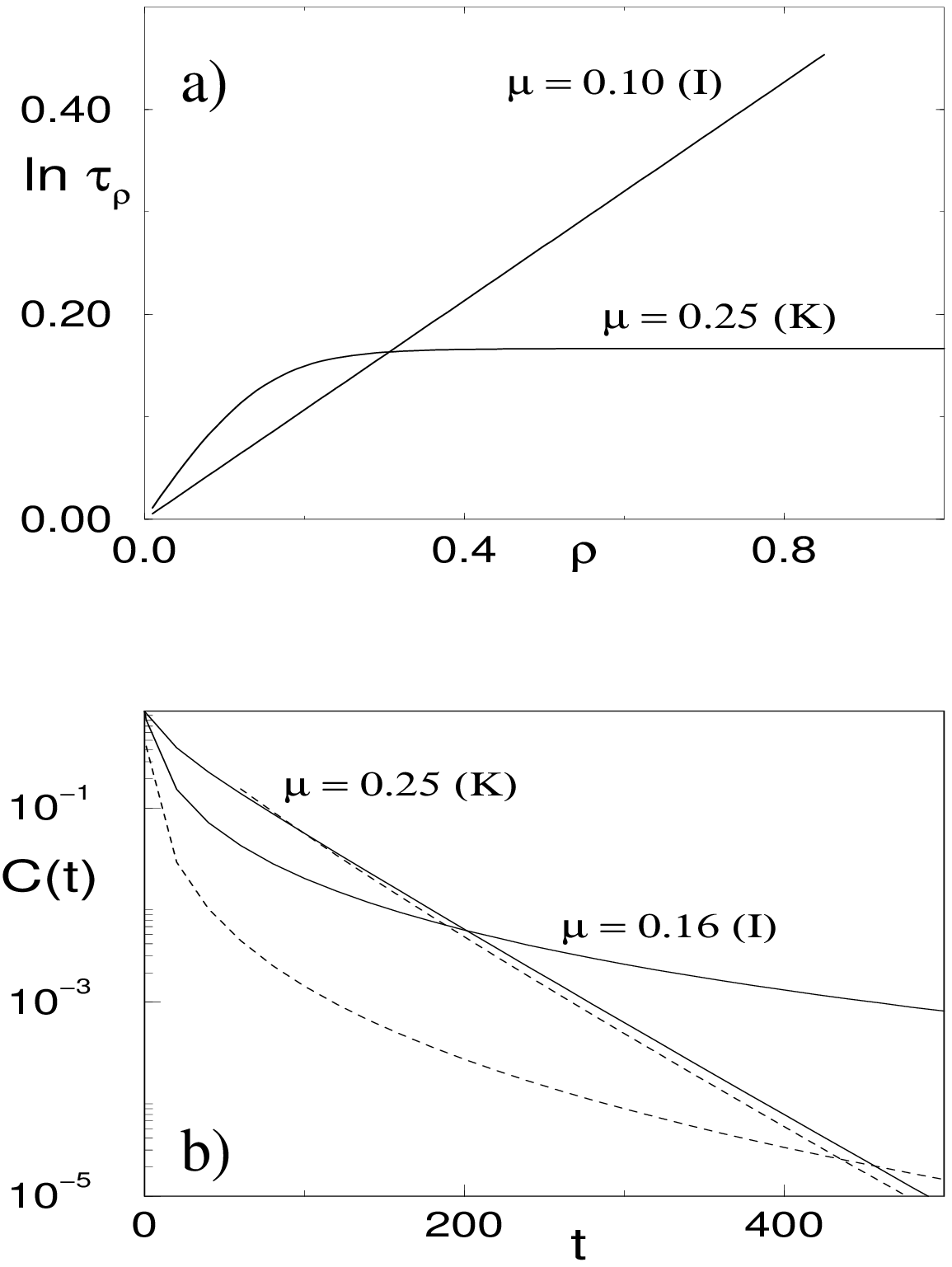}{(a) 
The time scales for relaxing out of different tilt states $\rho$ for the $I$ and $K$ models (scaled by $L$), 
for values of $\mu$ below the transition.  
(b) The tilt-tilt autocorrelation function for the two models. The full 
line is obtained from Eq.~\ref{corr2} while the dashed line is a result of the saddle point 
evaluation of Eq.~\ref{corr2}. $L$ was chosen to be $4096$ 
($I$) and $24$ ($K$) to make the time scales comparable. Time is  measured in units of $\Gamma_{o}^{-1}$.  
}{Tau_fig}  

The first consequence of the above form of the  transition rates  is that 
the time scale of relaxation out of a state with tilt
$\rho$, $\tau_{\rho} =
1/(W_{{\rho-1/L}, \rho} + W_{{\rho+1/L}, \rho})$, is a 
{\it non-decreasing} function of $\rho$. 
The exact expressions for $\tau_{\rho}$ (measured in units of 
$\Gamma_{o}^{-1}$), 
\be
{\tau_{\rho}}^{-1} = \left \{ \begin{array}{ll} 
	{ {e^{-{{16 \over 3} {\rho}\mu L}}+
e^{-L[{2 \over 3}{{\rm ln}2}+{2 \over 3}{{\rm ln}[\cos({\pi \over 3} 
- {{\pi {\rho}} \over 3})]}]}}} & ({I}) \\
	\\
	{{e^{-{2 \over 3}\mu L}+
e^{-L[{2 \over 3}{{\rm ln}2}+{2 \over 3}{{\rm ln}[\cos({\pi \over 3} 
- {{\pi {\rho}} \over 3})]}]}}} & ({K}) \\
\end{array} \right. ,   
\label{tauD1}
\ee
follow from Eqs.~\ref{barr1} and~\ref{barr2}, 
and they are plotted in  Fig.~\ref{Tau_fig}.
For the $I$ model the time scale increases
monotonically with $\rho$\cite{foot1}, as in the hierarchical models of Palmer {\em et al.} \cite{palmer}, while
in the $K$ model an initial rise is followed by 
saturation beyond $\rho_{min}$.   
Both are  in sharp contrast with Langevin  dynamics around the equilibrium 
state, for which the time to relax out of a given macro-state {\em decreases} the further the 
order parameter is away from its equilibrium value. (For example, in the Ising model, 
with Glauber dynamics and in the disordered phase, the  relaxation time out of 
a given magnetization state {\it decreases} with increasing magnetization.) 

The exponential separation between different $\tau_\rho$'s for 
all $\rho$ in the $I$ model, and for $\rho<\rho_{min}$ in the $K$ model, implies that the 
$\rho$ dynamics are {\em trap-like} \cite{bouchaud}. By this we mean that once the 
system decays out of state $\rho$ (in time $\tau_{\rho}$) it quickly looses memory of
where it came from.

To quantify the tilt dynamics  we compute the tilt-tilt autocorrelation function $C(t)$,  
defined as: 
\be
C(t) = {{\langle \rho(t) \rho(0) \rangle - {\langle \rho(0) \rangle}^{2}} \over 
{\langle {\rho(0)}^2 \rangle - {\langle \rho(0) \rangle}^{2}}},  
\label{corr1}
\ee 
with the average taken over different histories of $\rho$. 
Assuming trap-like dynamics, 
\be 
C(t) \approx {\sum_{\rho}{(\rho - {\langle \rho \rangle})}^2  e^{-F(\rho)} e^{-t/{\tau_{\rho}}} \over {\sum_{\rho}{(\rho - {\langle \rho \rangle})}^2  e^{-F(\rho)}}} ,
\label{corr2}
\ee
i.e., $C(t)$ is an equilibrium weighted average of relaxations out of different $\rho$ 
states. Such an approximate model works very well when checked against numerical 
simulations of the three-coloring model \cite{loops_epl}.  

The asymptotic decay of the autocorrelation functions obtained from
Eq.~\ref{corr2} can be extracted from a saddle point analysis of the sum and using a
quadratic approximation for the entropy (Eq.~\ref{entropyD1}). 
The autocorrelation functions obtained from the sum 
are compared to the saddle point solutions in  
Fig.~\ref{Tau_fig}b. The latter can be used
to analyze the asymptotic form of $C(t)$,  and the dependence of the relaxation time scales on $\mu$.  
In the limit of $t\to \infty$, and up to logarithmic corrections in $t$, we find:
\be
C(t) \sim \left \{ \begin{array}{ll} 
\exp\{-{3 \over 32}({{\mu_* - \mu} \over {\mu_*}^2})[{\rm ln}({{2 \mu_* t} \over {\mu_* - \mu}})]^2\} & (I) \\ 
\\
 \exp\{-{t \over {e^{2\mu L/3}}}~-~{{3 \sqrt{3}} \over {4 \pi}} 
[{\rm ln}{({t \over { e^{2 \mu L/3}}})}]^2\} & (K) \\
\end{array} \right. , 
\label{corrD1q} 
\ee
showing  that $C(t)$ has an exponential decay in the $K$
model, and a slower than exponential decay in  the $I$ model. From
Eq.~\ref{corrD1q} we also conclude that the relaxation times scale, $\tau$, for
the decay of $C(t)$ to an arbitrary constant $C_0$, diverges exponentially as
$\mu \rightarrow \mu_*$ for the $I$ model. This is the Vogel-Fulcher
type behavior observed in many fragile glass formers.  First order 
corrections to Eq.~\ref{corrD1q} lead to an even more rapid increase of time scales, 
with ${\tau}/{\rm ln}{\tau}$ diverging as Vogel-Fulcher.  In the $K$
model the corresponding time scales are exponential in $\mu$ (Arrhenius
form) with no divergence at the transition point.

\figone{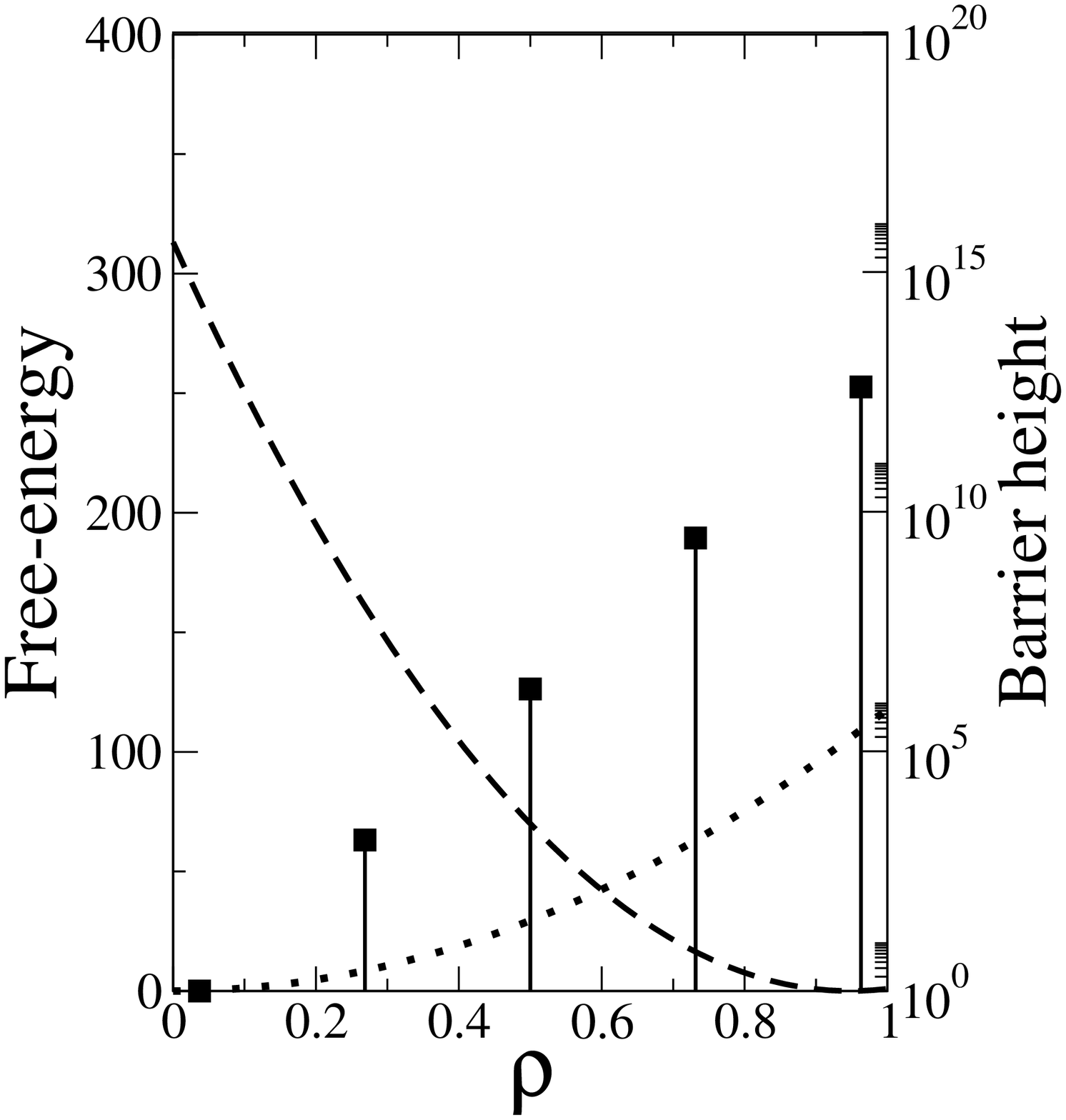}{Barrier height, $B(\rho)$ (dimensionless) shown as a set of solid
lines, and the quadratic approximation to the dimensionless free energies of the $I$ (dotted line) and $K$ (dashed
line) model; $\mu$ is chosen close to ${\mu}_*$ and  $L = 24$. Note the
logarithmic scale for the barrier height. }{barrier_fig}
 
The coarse grained dynamics defined by the transition matrix elements,
Eqs.~\ref{barr1} and~\ref{barr2}, were argued to follow from the
nonlocal loop dynamics of the dimer models. From this form of the $W$-matrix 
all the conclusions about critical dynamics of the $I$ and $K$ model are
derived.  We have confirmed this picture in considerable detail in
simulations of the  three
coloring model\cite{messina,Kolkata}, which, as discussed earlier,
is the loop weighted dimer model. The loop weights are not expected to 
affect the qualitative features of the energy and entropy functionals.  
Indeed, the measured $\tau_\rho$ for the $I$ and $K$ cases of the
three-coloring model compare very well\cite{messina,Kolkata} to the
analytical form plotted in Fig.~\ref{Tau_fig}. Similarly, the tilt-tilt
correlations show exponential decay in the $K$ model \cite{Kolkata} and
sub-exponential decay in the $I$ model \cite{messina,Kolkata} 
as expected from the analysis presented in this paper. The numerical evidence 
for Vogel-Fulcher type divergence of the relaxation time scale in the $I$
model was reported previously \cite{messina}.

The origin of the difference in the dynamical behavior of the $I$ and
$K$ models can be traced back to the interplay between the free energy
and the dynamical barriers.  The transition rates presented in
Eqs.~\ref{barr1} and \ref{barr2}, can be interpreted in terms of a
barrier $B({\rho}) = e^{(S(\rho-1/L)-S(\rho))}$ dividing the usual
Metropolis rates defined in terms of the free energy and leading to
Langevin dynamics \cite{palmer}.  These barriers increase
exponentially with $\rho$ as illustrated in Fig.~\ref{barrier_fig}.
Dynamics of the order parameter can be viewed as relaxation in the
free energy well in the presence of these barriers (see
Fig.~\ref{barrier_fig}).

In the $I$ model the free-energy minimum is
centered at $\rho = 0$ for all values of $\mu$. The well gets shallower as 
$\mu_*$ is approached implying that the system explores larger and
larger barriers leading to a divergent relaxation time scale.  
In the $K$ model, the minimum itself shifts to
larger values of $\rho$ and regions of increasing barrier heights 
(${\rho}_{min}\simeq 1$ in Fig.~\ref{barrier_fig}).
The dynamics is then completely dominated by a single time scale
determined by the barrier at $\rho_{min}$.

In conclusion, we have shown that classical dimer models with loop
dynamics are an interesting model system for studying glass-like
dynamical arrest close to a critical point. Our analysis shows that the
jammed nature of the dimer states plays a crucial role in determining
the nature of relaxation processes. To quantify how this is reflected
in the relaxation time scales we introduced an effective dynamical
model based on the master equation for the order parameter. We expect
that similar ideas can be carried over to the investigations of hard
sphere systems which are  of direct relevance to colloidal
experiments~\cite{weitz}.

Useful conversations with S.~Torquato are acknowledged. 
This work was supported by the NSF DMR-9815986 (DD, BC) and NSF DMR-9984471 (JK).

\end{multicols}


\begin{thebibliography}{0}

\bibitem{edwards}S. F. Edwards and D. V. Grinev, in {\it Jamming and Rheology: Constrained Dynamics on Microscopic and Macroscopic Scales}, eds A. J. Liu and S. R. Nagel, (Taylor and Francis, New York, 2001).
\bibitem{nagel_jamming}{ C.S.\ O'Hearn, S.A.\ Langer, A.J.\ Liu and S.R.\ Nagel, Phys. Rev. Lett. {\bf 86}, 111, (2001) }.
\bibitem{weitz} V.~Trappe {\em et al.}, Nature {\bf 411}, 772 (2001).
\bibitem{torquato}{S.~Torquato and F.H.~Stillinger, J. Phys. Chem. B {\bf 105}, 11849 (2001)}.
\bibitem{krauth}{L.~Santen and W.~Krauth, Nature {\bf 405}, 550 (2000)}.
%\bibitem{israeloff}{E.V.~Russel and N.E.~Israeloff, Nature {\bf 408}, 695 (2000)}.
%\bibitem{weitz}{E.R.~Weeks, J.C.~Crocker, A.C.~Levitt, A.~Schofield and D.A.~Weitz, Science {\bf 287}, 627 (2000)}. 
\bibitem{nagle_review}{J.F.~Nagle, C.S.O. Yokoi, and S.M. Bhattacharjee, in {\it Phase transitions 
and Critical Phenomena}, eds C. Domb and J.L. Lebowitz, Vol. 13, p. 235 (Academic Press, 1989).}
\bibitem{clh} C.L.~Henley, J. Stat. Phys. {\bf 89}, 483 (1997).  
\bibitem{halperin}P.C.~Hohenberg and B.I.~Halperin, Rev. Mod. Phys. {\bf 49}, 435 (1977).
\bibitem{angel_review}{C.A.~Angell, J. Phys. Chem. {\bf 49}, 863 (1988).}
\bibitem{kasteleyn}{P.W.~Kasteleyn, J. Math. Phys. {\bf 4}, 287 (1963)}.
\bibitem{blote}{H.W.J.~Bl\"ote and H.J.~Hilhorst, J.Phys.~A {\bf 15}, L631 (1982).}
\bibitem{foot2}The components of the tilt along the three lattice
directions of the triangular lattice are subject to two constraints.
One arising from the conservation of the total dimer number ($= L^2)$
and the other from periodic boundary conditions which force two of the
components to be equal. 
\bibitem{adhar}{A.~Dhar, P.~Chaudhuri and C.~Dasgupta, Phys. Rev. B {\bf 61},
6227 (2000).}
\bibitem{hui}{H. Yin and B. Chakraborty, Phys. Rev. Lett. {\bf 86},
2058 (2001), and  Phys. Rev. E {\bf 65}, 036119 (2002)}.
\bibitem{krauth1}{W.~Krauth and R.~Moessner, cond-mat/0206177}.
\bibitem{foot3}{In the $K$ model, only updates of $(0,1)$ loops involve an 
energy change.}
\bibitem{loops_epl}{D.~Das, J.~Kondev and B.~Chakraborty, cond-mat/0112281}.
\bibitem{foot1}Since the phase transition occurs along the metastable
branch in the $I$ model, the range of ${\rho}$'s participating in the
dynamics is restricted to ${\rho} \le {\rho}_{max}$. The free energy has a maximum at ${\rho}_{max}$ which 
vanishes as $\mu \rightarrow {\mu}_*$ and scales with the system size. 
\bibitem{palmer}{R.G. Palmer, D.L. Stein, E. Abrahams and
P.W. Anderson, Phys. Rev. Lett. {\bf 53}, 958 (1984)}; S. Teitel,
Phys. Rev. Lett. {\bf 60}, 1154 (1988).
\bibitem{bouchaud}{C. Monthus and J-P. Bouchaud, J.Phys. A {\bf 29}, 3847 (1996)}.
\bibitem{messina}{B. Chakraborty, D. Das and J. Kondev, European Physical Journal E, 2002 (to appear).}
\bibitem{Kolkata}{B. Chakraborty, D. Das and J. Kondev, Physica A, 2002 (to appear), and unpublished.}

\end{thebibliography}
\end{document}